\documentclass{aa}
\usepackage{latexsym}
\usepackage{times}
\usepackage{url}
\usepackage{graphicx}
\usepackage{natbib}

\bibpunct{(}{)}{;}{}{}{,}

\begin{document}

\thesaurus{07 % A&A Section 7
  ( 08.01.1; 08.01.3; 08.05.3; 08.09.2; 08.19.2; 13.21.5)}
  
\title{ORFEUS\,\textsc{II} echelle spectra: On the influence of iron-group  
  line blanketing in the Far-UV spectral range of hot subdwarfs}
 
\author{Jochen L.\ Deetjen}

\offprints{Jochen L.\ Deetjen}

\mail{deetjen@astro.uni-tuebingen.de}

\institute{Institut f\"ur Astronomie und Astrophysik, 
  Universit\"at T\"ubingen, 
  Waldh\"auser Str. 64, 
  D--72076 T\"ubingen, Germany 
  }

\date{Received 12 May 2000 / Accepted 14 June 2000}
 
\titlerunning{On the influence of the iron-group line blanketing in the
  Far-UV spectral range of hot subdwarfs}

\authorrunning{Jochen L.~Deetjen}

\maketitle

% --- abstract -------------------------------------------------------
%
\begin{abstract}
  We present an analysis of the subdwarf O~star Feige~67 with a fully
  metal-line blanketed NLTE model atmosphere based on high-resolution
  Far-UV (912--2000\,\AA) and Near-UV (2000--3400\,\AA) spectra from new
  ORFEUS\,\textsc{ii} echelle observations and the IUE final archive.  The  
  Far-UV spectra are heavily blanketed by iron and nickel lines, preventing 
  the detection of the stellar continuum and complicating the abundance
  analysis.
  
  Important points concerning the account for blanketing by millions of
  iron-group lines and for an accurate determination of iron and nickel
  abundances are discussed: The usage of all theoretically and
  experimentally known line opacities of the iron-group elements, the
  consideration of a broad wavelength range for a reliable determination of
  the stellar continuum flux, and the role of interstellar reddening. This
  paper outlines a basic approach for spectral analysis of future
  high-resolution Far-UV spectra of hot compact stars. During this study we
  re-analyzed the iron and nickel abundance of our exemplary object
  Feige~67 and confirm their extraordinarily amount.
  
\end{abstract}

% === BODY ===========================================================
%

% --- figure 1 -------------------------------------------------------
%
\begin{figure*}
  \resizebox{\hsize}{!}{\includegraphics{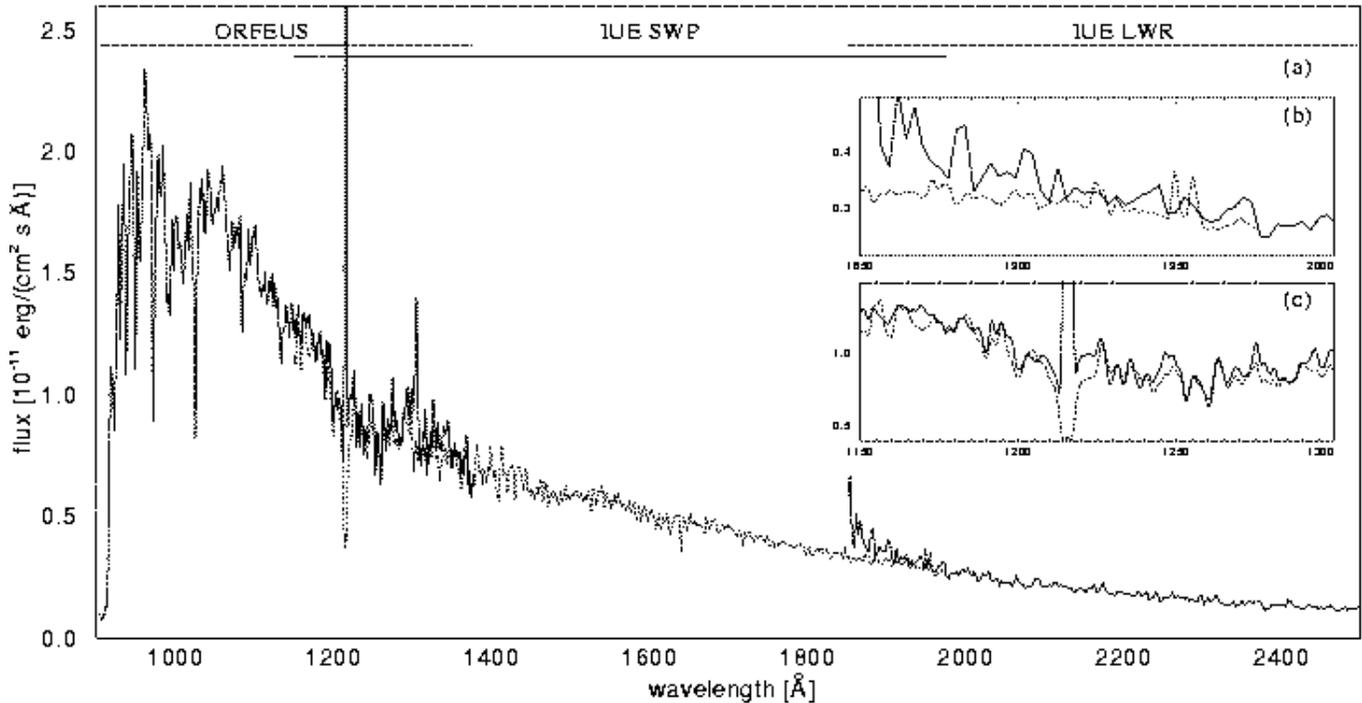}}
  \caption{The combined ORFEUS\,\textsc{ii} and IUE spectra of
    Feige~67. The small inset figures show in more detail the overlap
    regions of the three spectra, revealing calibration problems at the
    blue end of the IUE-LWR spectrum. The spectra are smoothed with a
    Gaussian of FWHM\,=\,2.0\,\AA}
  \label{fig:LWP_SWP_ORF_flux-calibration}
\end{figure*} 
% 
% --- figure 2 -------------------------------------------------------
%
\begin{figure*}
  \resizebox{\hsize}{!}{\includegraphics{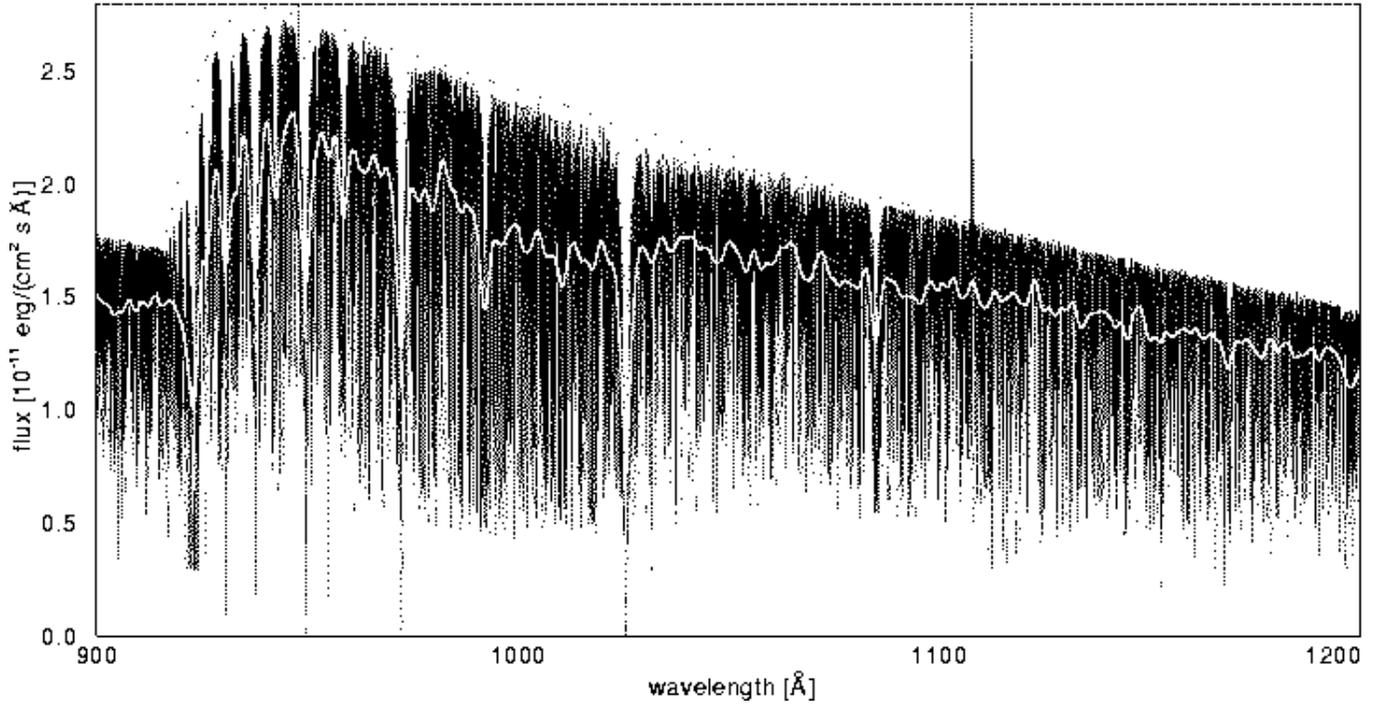}}
  \caption{Synthetic spectrum of the final model (including interstellar
    Lyman line absorption) for Feige 67, demonstrating the high density of
    Fe and Ni lines. Overplotted (white line) is the same spectrum, but
    degraded with a Gaussian of FWHM\,=\,2.0\,\AA, matching the smoothing
    of the observations shown in
    Fig.~\ref{fig:LWP_SWP_ORF_flux-calibration}. The dash-dotted line
    indicates the continuum model flux. Model parameters are given in
    Tab.~\ref{tab:properties}} 
  \label{fig:unsmoothed-smoothed}
\end{figure*}  
%
% --- figure 3 -------------------------------------------------------
%
\begin{figure*}
  \resizebox{\hsize}{!}{\includegraphics{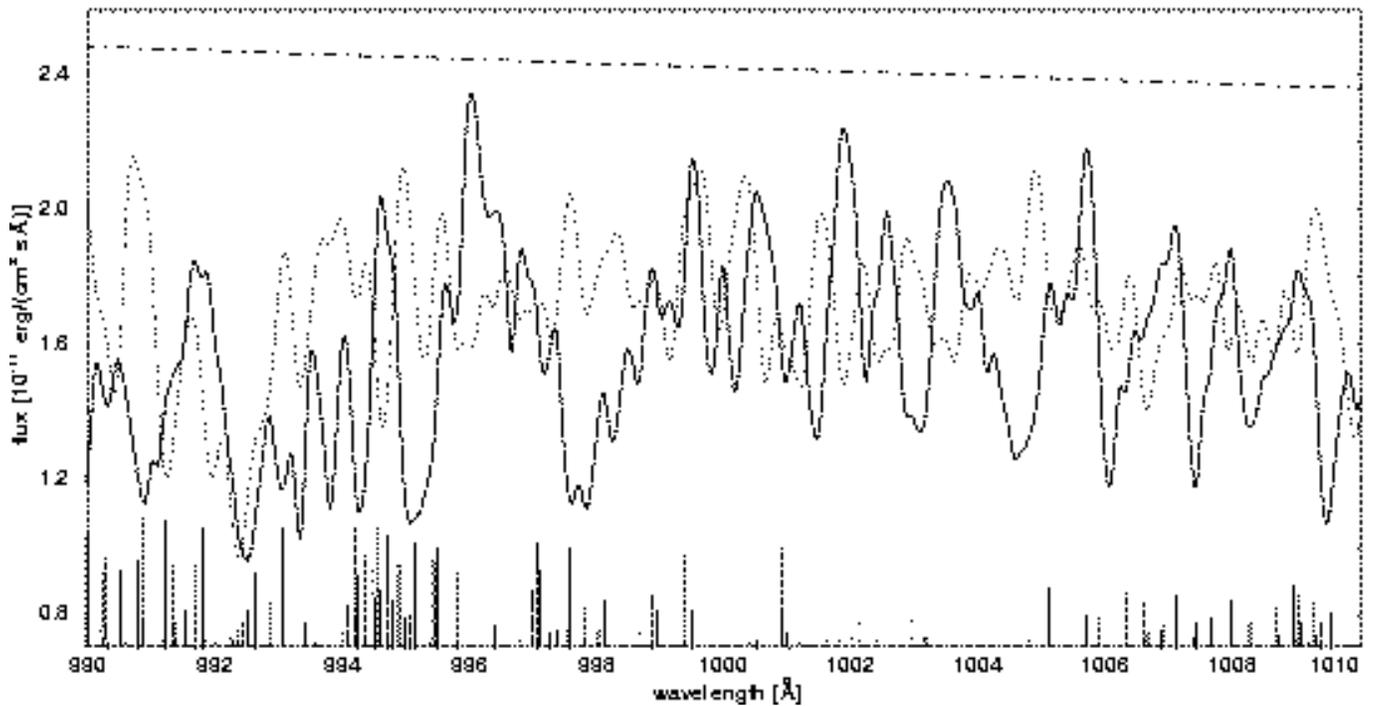}} 
  \caption{Detail of the ORFEUS\,\textsc{ii} spectrum (thick line) with the   
    final model (thin line). The flux level, which falls well below the 
    continuum (dash-dotted line) is matched satisfactorily, however, the
    identification of individual lines is not possible because of
    inaccurate line positions in the full Kurucz line list. Vertical lines 
    at the bottom panel mark Ni\,/\,Fe lines in the model with gf-values
    $\ge -2.0$ (bar length proportional to log\,gf). The spectra are
    smoothed with a Gaussian of FWHM\,=\,0.25\,\AA}
  \label{fig:lines-990-1010}
\end{figure*}  
%
% --- figure 4 -------------------------------------------------------
%
\begin{figure*}
  \resizebox{\hsize}{!}{\includegraphics{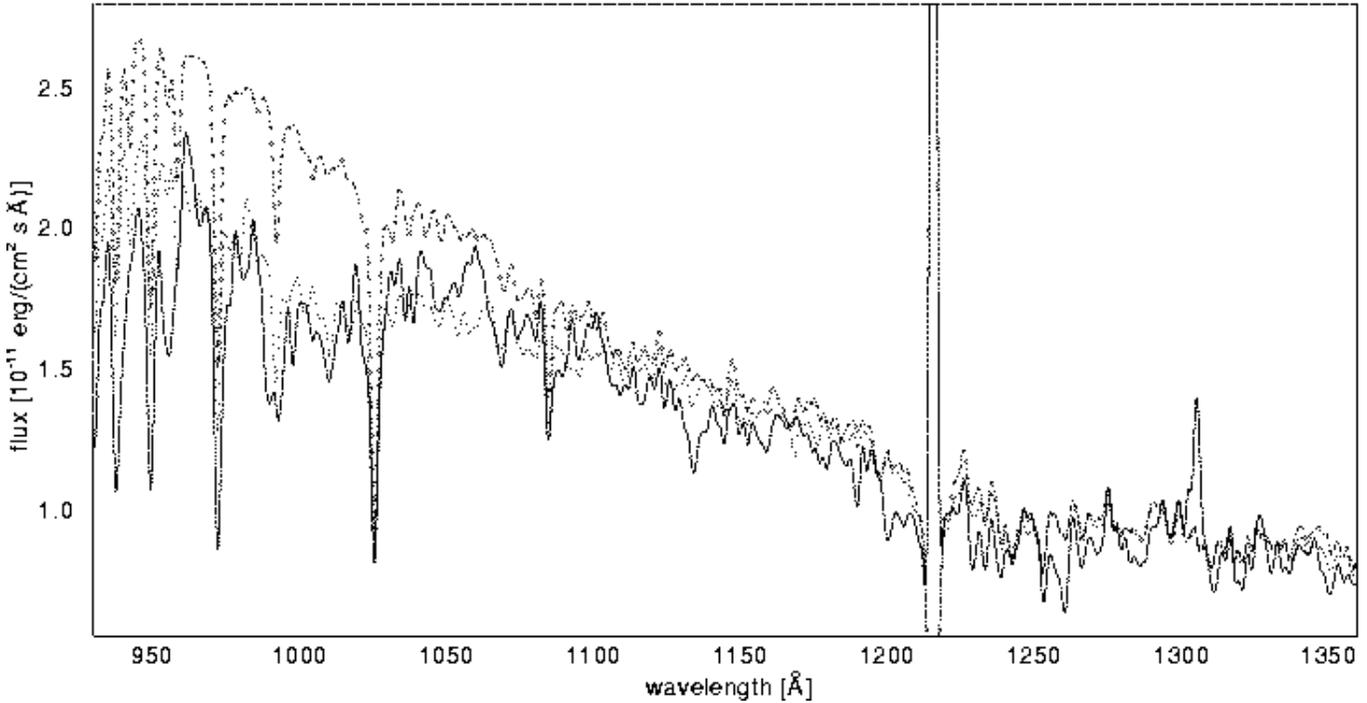}} 
  \caption{ORFEUS\,\textsc{ii} spectrum (thick line) compared to two model 
    spectra (including reddening and H\textsc{i} absorption) which are
    calculated with the full Kurucz line list in one case (thin line) and
    the small subset of experimentally known lines in the other case (thin
    line with circles). The full list is necessary to reproduce the
    spectral shape particularly below 1100\,\AA. All spectra are smoothed
    with a Gaussian of FWHM\,=\,2.0\,\AA}
  \label{fig:large-small}
\end{figure*}  
%
% --- figure 5 -------------------------------------------------------
%
\begin{figure*}
  \resizebox{\hsize}{!}{\includegraphics{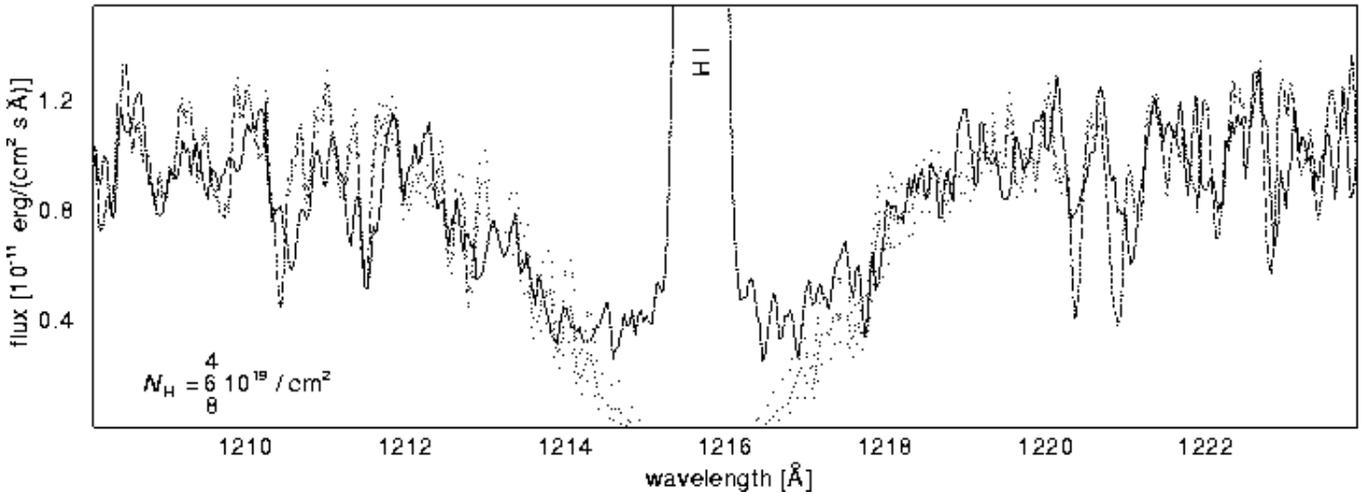}}
  \caption{The interstellar H\,\textsc{i} column density is derived from
    this fit to the Ly$\alpha$ profile in the ORFEUS\,\textsc{ii}
    spectrum. The spectra are smoothed with a Savitzky-Golay filter
    (observation: M=4, n$_L$+n$_R$+1=9; model: M=4, n$_L$+n$_R$+1=17)}
  \label{fig:nh_orfeus}
\end{figure*}  
%
% --- figure 6 -------------------------------------------------------
%
\begin{figure*}
  \resizebox{\hsize}{!}{\includegraphics{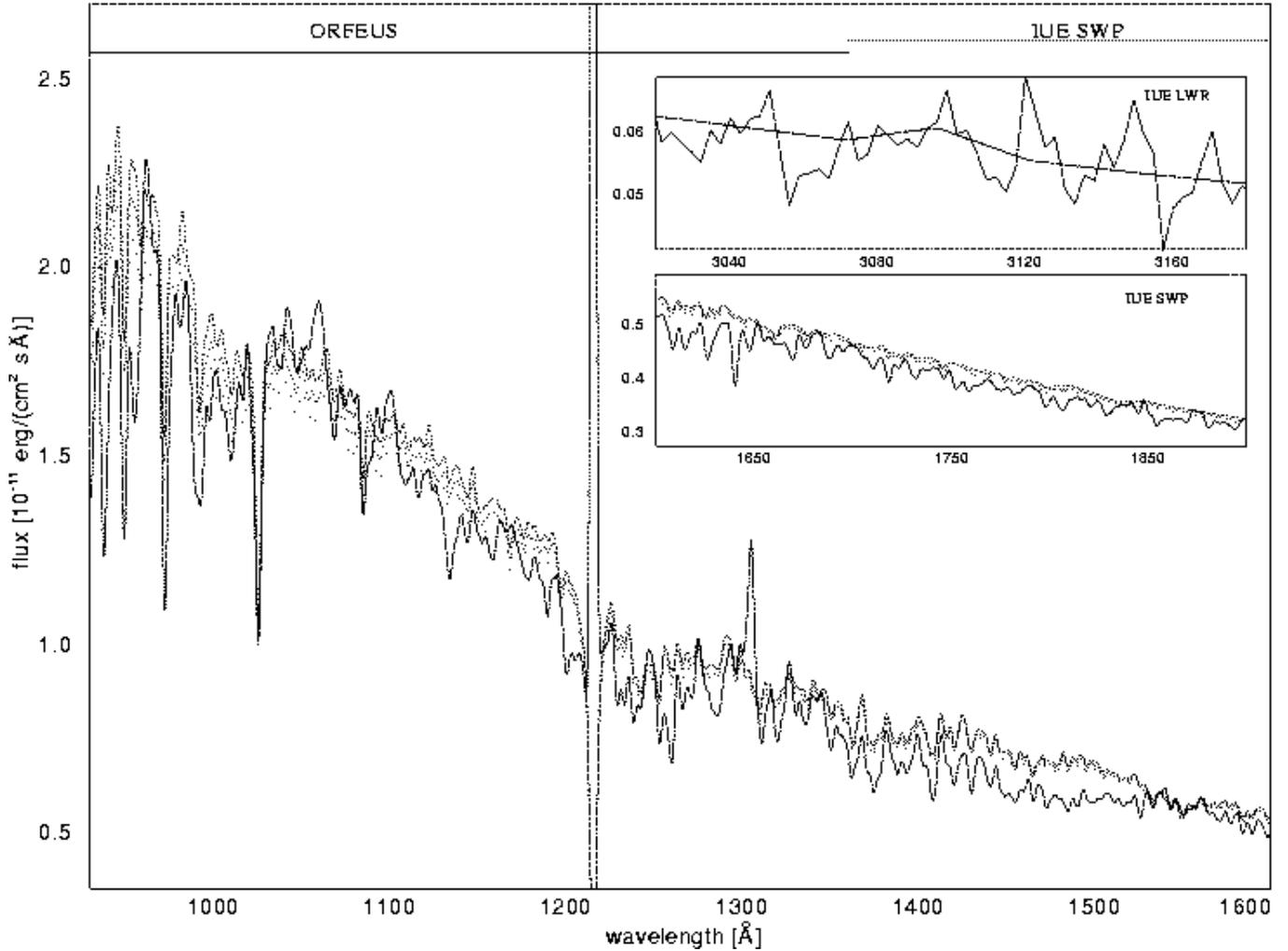}}
  \caption{ORFEUS\,\textsc{ii} spectrum (thick line) compared to the final 
    model which is not attenuated by the interstellar extinction (thin
    dash-dotted line) and to final model which is attenuated by
    interstellar extinction for two values of E(B-V), representing the
    error margin (E(B-V)=0.011 and 0.021). The inset figures show the IUE
    spectra for comparison. The spectra are smoothed with a Gaussian of
    FWHM\,=\,2.0\,\AA and the reddened models are normalized to the
  observation near 3000\,\AA}
  \label{fig:ebv-variation}
\end{figure*}  
%
% --- figure 7 -------------------------------------------------------
%
\begin{figure*}
  \resizebox{\hsize}{!}{\includegraphics{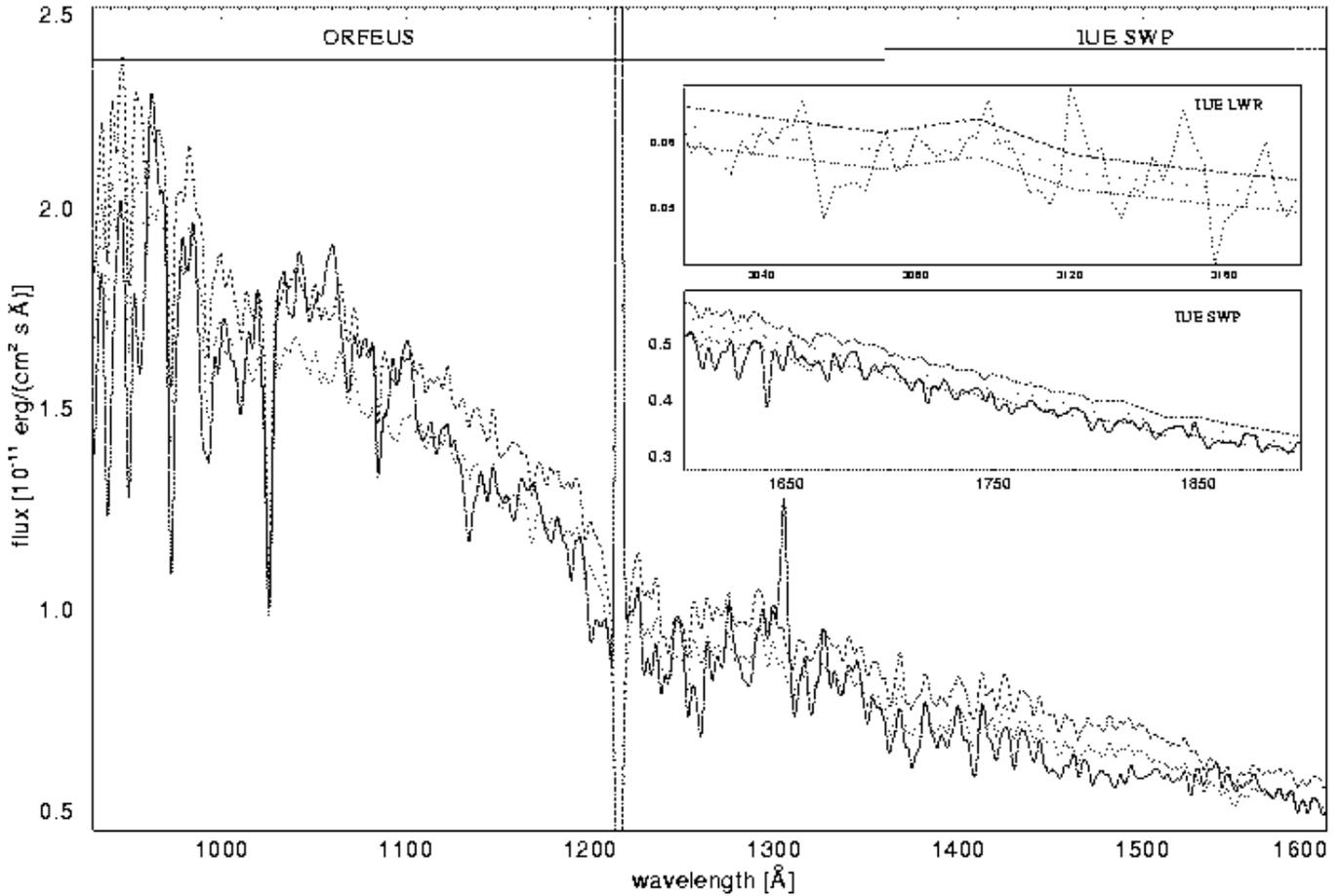}}
  \caption{The synthetic spectrum is normalized to the observation (thick 
    line) near $\lambda$\,=\,3000\,\AA\ to avoid the strong line blanketing
    in the Far-UV.  The two thin lines denote the normalized model spectrum 
    with the normalization factor varied by $\pm$5\%, representing the
    uncertainty in normalization. The two inset figures show details from
    the IUE spectra with the best fit model (thin dashed-dotted line)
    plotted additionally. All spectra are smoothed with a Gaussian of
    FWHM\,=\,3.0\,\AA}
\label{fig:scalingfactor}
\end{figure*}  
%
% --- figure 8 -------------------------------------------------------
%
\begin{figure*}
  \resizebox{\hsize}{!}{\includegraphics{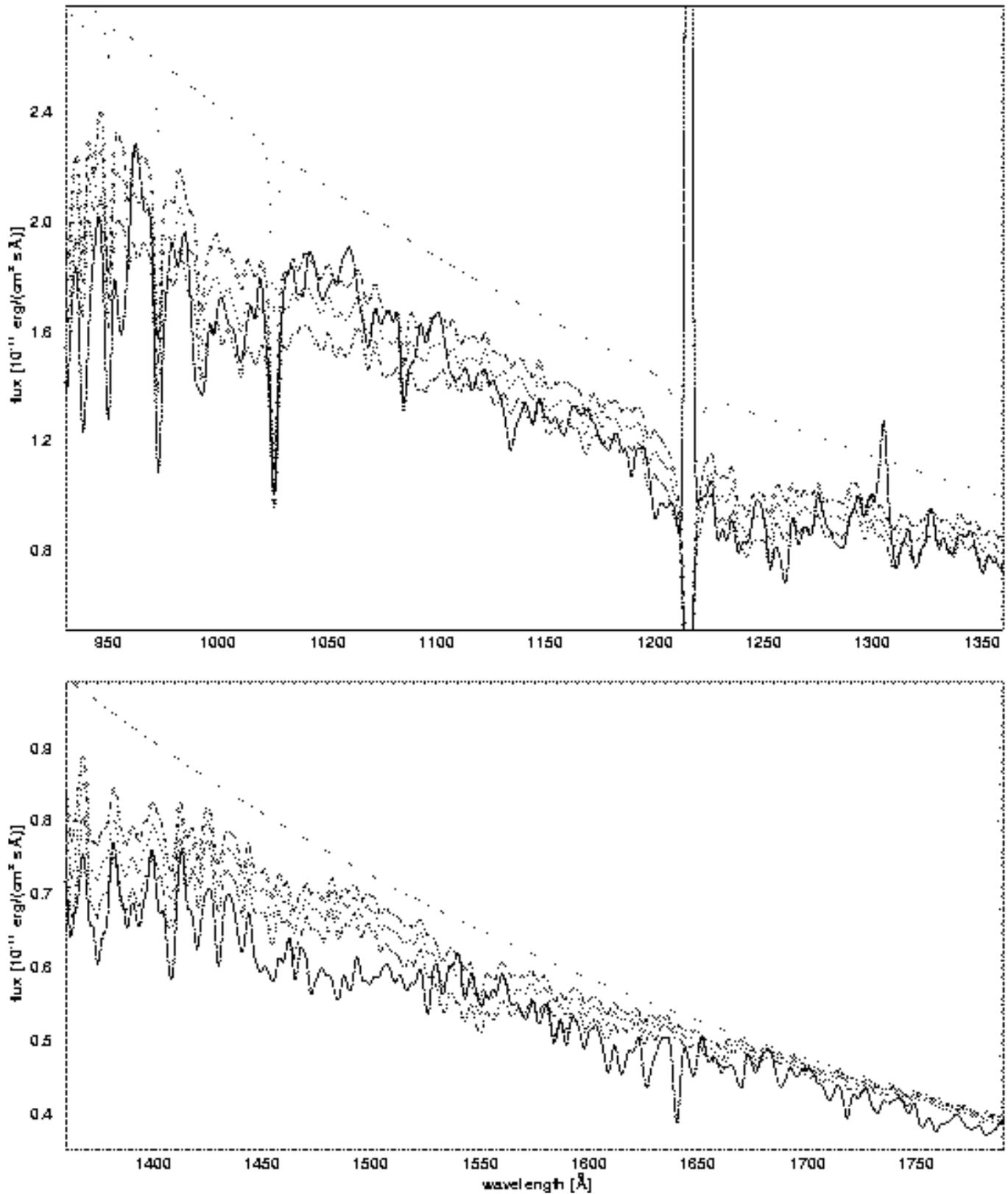}}
  \caption{Detailed comparison of the final model (thin line) to the
    ORFEUS\,\textsc{ii} (top panel) and IUE-SWP (bottom panel) spectra
    (thick lines). The other two thin lines represent models with Fe and Ni 
    abundances varied by a factor two, representing the analysis
    error. Between 1500\,\AA\ and 1600\,\AA\ all models overestimate the
    stellar flux, perhaps due to missing opacity from other iron group
    elements which are disregarded in this study. The dashed-dotted line
    denotes the model continuum. All spectra smoothed with a Gaussian of
    FWHM\,=\,3.0\,\AA}
  \label{fig:abu-variation-ORF+IUE}
\end{figure*}  
%
% --- figure 9 -------------------------------------------------------
%
\begin{figure*}
  \resizebox{\hsize}{!}{\includegraphics{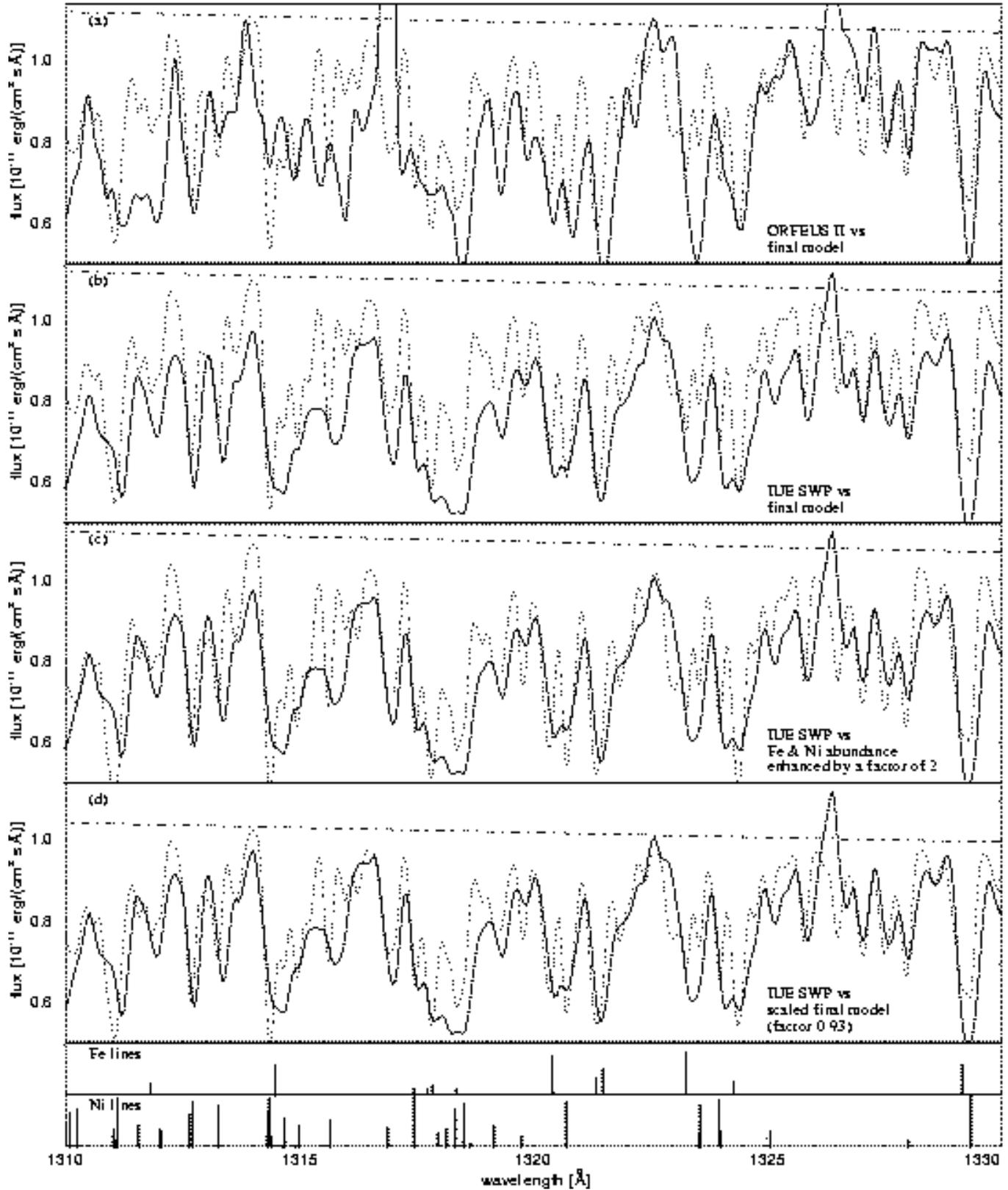}}
  \caption{Detail of another UV wavelength region of Feige~67, where IUE
    and ORFEUS\,\textsc{ii} spectra overlap, showing observed spectra
    (thick line) and final model spectra (thin line). The dashed-dotted
    line represents the model continuum. Panel (a): ORFEUS\,\textsc{ii}
    vs.\ final model. Panel (b): IUE-SWP vs. final model. Panel (c):
    IUE-SWP vs. model with Fe and Ni abundances enhanced by a factor
    two. Panel (d): IUE vs. scaled model (factor 0.93). All spectra are
    smoothed with a Gaussian of FWHM\,=\,0.2\,\AA }
  \label{fig:lines-1300-1330}
\end{figure*}  
%
% --------------------------------------------------------------------

% ----- TABLE #1 -----------------------------------------------------
%
\begin{center}
\begin{table}
  \caption{Observation log and properties of
    \object{Feige~67}. Photospheric parameters are taken from
    \cite{haas:phd}, except for the Fe and Ni abundances which have been
    re-analyzed in this paper} 
  \label{tab:properties}
  \begin{tabular}{l l}
    \hline
    \noalign{\smallskip}
    Alias
    &  \object{BD+18~2647} \\ 

    m$_\textrm{V}$ 
    & 12.1 \\

    \noalign{\medskip}

    ORFEUS\,\textsc{ii} IDs 
    & 5256\_1, 5256\_2 \\
    
    Observation\ date      
    & 01.12.\,/\,03.12. 1996 \\
    
    Exposure\ time [s]     
    & 2954 \\
    
    \noalign{\medskip}

    IUE LWR IDs (low res.)
    & 03400, 01570, 02944 \\

    Observation\ date      
    & 27.05.\,/\,16.11. 1978, 06.01. 1979 \\
    
    Exposure\ time [s]     
    & 240, 480, 597 \\
    
    \noalign{\medskip}

    IUE SWP ID (high res.)
    & 20488 \\
 
    Observation\ date      
    & 20.07. 1983 \\
    
    Exposure\ time [s]     
    & 10\,800 \\
    
    \noalign{\medskip}

    n$_\textrm{H \textsc{i}}$ column density
    & $(6\pm1) \cdot 10^{19}\,/\textrm{cm}^{2}$ \\
    
    Color excess E$_{\textrm{B-V}}$ 
    & $0.016 \pm 0.005$ \\

    \noalign{\medskip}

    $T_{\rm eff}$ [K]   
    & 60\,000 $\pm$ 4\,000 \\
    
    $\log g$ [cm/s$^2$]       
    & 5.0 \\
    
    \noalign{\medskip}

    n$_{\rm He}$/n$_{\rm H}$ 
    & $5.3 \cdot 10^{-2}$ \\

    n$_{\rm C}$/n$_{\rm H}$ 
    & $3.2 \cdot 10^{-5}$ \\

    n$_{\rm N}$/n$_{\rm H}$ 
    & $1.4 \cdot 10^{-5}$ \\

    n$_{\rm O}$/n$_{\rm H}$ 
    & $7.3 \cdot 10^{-5}$ \\

    n$_{\rm Fe}$/n$_{\rm H}$ 
    & $1.4 \cdot 10^{-4}$ \\

    n$_{\rm Ni}$/n$_{\rm H}$ 
    & $4.2 \cdot 10^{-5}$ \\

    \noalign{\smallskip}
    \hline
  \end{tabular}
\end{table}
\end{center}
%
% --------------------------------------------------------------------

% ----- TABLE #2 -----------------------------------------------------
%
\begin{table}
  \caption{
    Summary of model atoms used in our model atmosphere
    calculations. Numbers in brackets denote individual levels and lines
    used in the statistical NLTE line blanketing approach for iron and
    nickel. The model atom for each chemical element is closed by a single
    level representing the highest ionization stage (not listed
    explicitly)}  
  \label{tab:atoms}

  \begin{tabular}{l l r r r r}
    \hline
    \noalign{\smallskip}
    element & ion & \multicolumn{2}{l}{NLTE levels} & lines & \\
    \noalign{\smallskip}
    \hline
    \noalign{\smallskip}
    H  & \textsc{i}   &  16 & &  56 & \\
    \noalign{\smallskip}
    He & \textsc{i}   &  29 & &  91 & \\
       & \textsc{ii}  &  32 & & 115 & \\
    \noalign{\smallskip}
    C  & \textsc{iii} &  58 & & 231 & \\
       & \textsc{iv}  &  57 & & 201 & \\
    \noalign{\smallskip}
    N  & \textsc{iii} &   1 & &   0 & \\
       & \textsc{iv}  &  90 & & 368 & \\
       & \textsc{v}   &  36 & &  98 & \\
    \noalign{\smallskip}
    O  & \textsc{iv}  &  30 & &  99 & \\
       & \textsc{v}   &  44 & &  86 & \\
       & \textsc{vi}  &  52 & & 237 & \\
    \noalign{\smallskip}
    Fe & \textsc{iv}   & 7 & (6\,472) & 25 & (1\,027\,793)  \\
       & \textsc{v}    & 7 & (6\,179) & 25 &    (793\,718)  \\
       & \textsc{vi}   & 8 & (3\,137) & 33 &    (340\,132)  \\
       & \textsc{vii}  & 9 & (1\,195) & 39 &     (86\,504)  \\
       & \textsc{viii} & 7 &    (310) & 27 &      (8\,724)  \\
    \noalign{\smallskip}
    Ni & \textsc{iv}   & 7 & (5\,514) & 25 &    (949\,506) \\
       & \textsc{v}    & 7 & (5\,960) & 22 & (1\,006\,189) \\
       & \textsc{vi}   & 7 & (9\,988) & 22 & (1\,110\,584) \\
       & \textsc{vii}  & 7 & (6\,686) & 18 &    (688\,355) \\
       & \textsc{viii} & 7 & (3\,600) & 27 &    (553\,549) \\
    \noalign{\smallskip}
    \hline
    \noalign{\smallskip}
    total   &     & 525 & (49\,041) & 1845 & (6\,565\,054)\\
    \noalign{\smallskip}
    \hline
  \end{tabular}
\end{table}
%
% --------------------------------------------------------------------

% --- section: Introduction ------------------------------------------
%
\section{Introduction} 

With the launch of FUSE in 1999 the window to high-resolution Far-UV
spectroscopy has been opened again. In order to prepare an adequate
analysis of the upcoming data for hot compact stars, we studied already 
available high-resolution ORFEUS\,\textsc{ii} echelle spectra using
state-of-the-art NLTE (non-local thermodynamic equilibrium) model
atmospheres with elaborated model atoms including the opacities of millions 
of iron and nickel lines.

This paper pursues three aims: First we want to outline a basic approach
for spectral analysis of future high-resolution Far-UV spectra under the
viewpoint of iron-group line blanketing. The second aim is a presentation
of possible error sources and an estimation of the uncertainties of the
abundances derived.  The third aim is an improved determination of the iron
and nickel abundance of our demonstration object, the sdO Feige~67.

A commonly used approach to determine Fe and Ni abundances is the analysis
of individual, prominent line profiles within a narrow wavelength range.
For this purpose the observed and the synthetic spectra are normalized to
the continuum.  However, we will show that the detection of the stellar
continuum can be extremely difficult or impossible because of the presence
of millions of lines from the iron-group elements. Hence, derived
abundances may be afflicted with substantial errors. We will demonstrate
that reliable analyses of Far-UV spectra, like from e.g.\ ORFEUS or FUSE,
require a sufficiently large spectral base into the Near-UV region, where
line blanketing is less severe, so that the model flux can be normalized to
the continuum in that range. For this purpose we analyzed the combined
ORFEUS\,\textsc{ii} and IUE spectra of the sdO Feige~67, covering the
wavelength range from 900\,\AA~ to 3200\,\AA.

Feige~67 is well suited to demonstrate the difficulties and possible error 
sources related to the Fe and Ni line blanketing effect, because it is
known for its overabundance of the elements (probably due to radiative
levitation) and because high-resolution spectra are available and the
stellar parameters (see Tab.~\ref{tab:properties}) are well known
\citep{haas:phd, werner:98a}. \cite{haas:phd} derived the effective
temperature by the ionization equilibria of Fe and Ni.  Our re-analysis of
the Fe and Ni abundances results in smaller values compared to the
abundances derived by \cite{haas:phd}, however, we expect no strong
back-reaction on the atmospheric structure which in turn might lead to a
different $T_{\rm eff}$. Thus we concentrate here on a thorough analysis of 
the effect of different Fe and Ni abundances on the synthetic spectrum.

In the following we describe briefly the observation and data reduction
(Sect.~2) followed by a short outline of the model atmospheres used
(Sect.~3).  In Sect.~4 we discuss the four most important steps of the
analysis and present the major error sources by means of illustrative
figures. The results are summarized in Sect.~5.

% --- section: Observation and data reduction ------------------------
%
\section{Observations}

The echelle spectrometer was one of the focal instruments of the ORFEUS
telescope, which was flown on its second mission in November\,/\,December
1996.  The wavelength range is 900-1400\,\AA\ with a spectral resolution of
$\lambda/\Delta\lambda = 10\,000$. Two separate observations of Feige~67
were obtained with a total integration time of 2954\,s (see
Tab.~\ref{tab:properties}). The two echelle images were co-added and then
the standard extraction procedure was applied \citep{barnstedt:99a}.
Feige~67 has been observed with the IUE satellite in the short and in the
long wavelength range (see Tab.~\ref{tab:properties}). For our analysis we
used the so-called ``preview spectra'' available in the IUE final archive.

Any relative error in the flux calibration of the different instruments
with respect to each other will propagate into the error bar of the stellar
parameters to be derived, because we will rely on the correct overall shape
of the combined spectrum. Fig.~\ref{fig:LWP_SWP_ORF_flux-calibration} shows
the good flux calibration of all three observations of Feige~67, because of
a good match of the adjoining spectral regions. On hand of our final model
we show in Fig.~\ref{fig:unsmoothed-smoothed} the strong line blanketing in
the Far-UV spectral region, which results in an overall flux depression.
particularly in the degraded spectrum.

% --- section: NLTE Model atmospheres -------------------------------------
%
\section{NLTE Model atmospheres}

Based on the results of \cite{haas:phd} we calculated a plane-parallel
non-LTE model atmosphere with \mbox{$T_{\rm eff} = 60\,000$} and log\,g =
5.0 in radiative and hydrostatic equilibrium. According to \cite{haas:phd}
the Fe and Ni abundances were set to 10 and 70 times solar, respectively.
In order to calculate synthetic spectra with different Fe and Ni
abundances, a single formal solution of the radiation transfer equation is
performed keeping fixed the whole model structure.

The computer code is based on the Accelerated Lambda Iteration method
\citep{werner:85a,werner:86a} and it can handle the line blanketing of
iron-group elements by a statistical approach using superlevels and
superlines with an opacity sampling technique \citep{anderson:85a,
  werner:99a} as described in \cite{deetjen:99a}.

A summary of the detailed model atoms used is given in
Tab.~\ref{tab:atoms}.  Test calculations for Feige~67 have shown, that
\ion{Fe}{v} and \ion{Fe}{vi} are the dominant ionization stages in those
atmospheric layers where the iron lines are formed.

For the calculation of opacities of iron group elements one can choose
between a subset of the Kurucz \citep{kurucz:91a} line list, containing
only lines known from laboratory spectra, and the full one, augmented by a
vast number of theoretically computed lines. The full line list is
afflicted with uncertainties in oscillator strengths and wavelength
positions, preventing the identification of individual lines when comparing
with observed spectra (see Fig.~\ref{fig:lines-990-1010}). The small list
on the other hand causes much too less total opacity.
Fig.~\ref{fig:large-small} demonstrates the high relevance of using the
full line list for further analysis in the Far-UV range.  It also
demonstrates the impossibility to determine the stellar continuum, so that
analyses with model atmospheres must rely on a correct flux normalization
in the Near-UV where line blanketing is less severe.

% --- section: Analysis ----------------------------------------------
%
\section{Analysis}

In the following we discuss and determine the most important parameters
which influence the observed spectrum. Due to their interacting character
the analysis has to be done iteratively.

% --- subsection ---
%
\subsection{H~\textsc{i} column density}
The interstellar H\textsc{i} column density along the line of sight is
derived from the Ly$\alpha$ profile in the ORFEUS\,\textsc{ii} spectrum
(Fig.~\ref{fig:nh_orfeus}).  We applied different models and found no
dependency of the derived column density on the adopted Fe and Ni abundance
(which might be caused by strong blending of the Ly$\alpha$ profile) and
obtained N$_{\textrm{H}\textsc{i}} = (6\pm1) \cdot 10^{19} /\textrm{cm}^2$.
The same result has been derived from the Ly$\alpha$ profile in the IUE-SWP
spectrum. Note that we did not try to normalize the model continuum to a
putative local stellar continuum near Ly$\alpha$, instead we relied on the
flux normalization in the Near-UV (see below).  Therefore interstellar
reddening has to be accounted for, otherwise one underestimates
N$_{\textrm{H}\textsc{i}}$ by approximately 0.1\,dex. This in turn (as
reddening is calculated from N$_{\textrm{H}\textsc{i}}$) would cause an
overestimation of the iron-group abundance by about a factor two.

Another source of error can be too strong smoothing of the observation or
the model. Even convolution with a Gaussian of only FWHM\,=\,0.2\,\AA~
corrupts the interstellar Ly$\alpha$ profile because of many blending iron
group lines.  Therefore we used a low-pass filter for noise-reduction, the
Savitzky-Golay smoothing filter, as described in \cite{recipes:95}.

% --- subsection ---
%
\subsection{Interstellar reddening}
Based on the determined column density, the color excess can be calculated 
to E$_{(B-V)} = 0.016 \pm 0.005$, using the formula of
\cite{groenewegen:89a}
\begin{displaymath}
N_{\textrm{H}\textsc{i}} = (3.8 \pm 0.9) \times 10^{21} \: E_{(B-V)}.
\end{displaymath}
An independent determination of the color excess could not be realized,
because it is too small to produce a detectable 2200\,\AA~ feature in the
IUE-LWR spectrum.

In the literature the applicability of the above formula is debated as well
as the choice of the interstellar extinction law itself. Hence some authors
try to determine such a law adapted to their specific observation.
Generally, however, this is impossible, so that in our case we expect a
systematic error of at least 0.1\,dex (using the interstellar extinction
law as described in \cite{seaton:79a}) in addition to the statistical
error, which is estimated to 0.1\,dex from Fig.~\ref{fig:ebv-variation}.
It is clear that the neglect of interstellar reddening in the Far-UV range
results in an underestimation of the stellar flux level. This would lead to
an overestimation of the Fe and Ni abundances, by about a factor of
1.0\,dex in our specific case.

% --- subsection ---
%
\subsection{Normalization of the model to the observation}
As already mentioned above, strong line blanketing in the Far-UV requires a
model flux normalization in the Near-UV, where reduced blanketing does not
prevent detection of the stellar continuum. We determined the position of
the continuum at the red end of the spectrum above 3000\,\AA. In
Fig.~\ref{fig:scalingfactor} the resulting synthetic flux has been
multiplied by a factor of 0.95 and 1.05, respectively, representing the
uncertainty of fixing the continuum in the Near-UV. A determination of the
continuum only within the Far-UV range results in an analytical uncertainty
in the iron and nickel abundance of about a factor two (compare
Fig.~\ref{fig:scalingfactor} with \ref{fig:abu-variation-ORF+IUE}).
Neglecting the general slope of the spectrum in the Near-UV and the Far-UV
may result an additional systematic error of the same magnitude.

% --- subsection ---
%
\subsection{Variation of the iron and nickel abundance}
After the preparatory steps discussed above one is now in the position to
determine the abundances of Fe an Ni by fitting the spectral slope.
Fig.~\ref{fig:abu-variation-ORF+IUE} demonstrates the sensitivity of the
flux distribution against the simultaneous variation of the Fe and Ni
abundances in the model by $\pm 0.3$\,dex. Detailed calculations have
shown, that modifications in the Fe abundance are mainly noticeable in the
ranges 960--1140\,\AA~ and 1360--1510\,\AA, whereas modifications in the Ni
abundance occur dominantly in the range 1030--1330\,\AA. The resulting
abundances of the best fitting model are given in
Tab.~\ref{tab:properties}.

A determination of the Fe and Ni abundance by fitting the profile of
individual lines (Fig.~\ref{fig:lines-1300-1330}) is rather problematic
because of, as stated above, blending by lines with inaccurately known
wavelength positions and oscillator strengths.

% --- section: Conclusion --------------------------------------------
%
\section{Conclusion}

In order to derive the iron and nickel abundances in the sdO star Feige~67
we proceeded in two major steps. At first we considered the shape of the
spectrum over a broad wavelength range.  Here we focused on the correct
consideration of all effects of physical relevance, i.e.\ large model
atoms, n$_\textrm{H \textsc{i}}$ determination, and interstellar reddening
as well as on effects of technical relevance, i.e. determination of the
continuum and quality of the flux calibration. As a result we could
demonstrate the important influence of these details for the second step,
where the Fe and Ni abundances are derived from the ORFEUS\,\textsc{ii}
spectrum.

We confirmed the high Fe and Ni abundances of Feige~67 and find 3.5 and
26.5 times solar, respectively (see Tab.~\ref{tab:properties}), which are
however a factor three smaller than the results by \cite{haas:phd}.
\cite{becker:95a, becker:95c, becker:95b} derived Fe and Ni abundances of
approximately 10 times solar, based on the IUE-SWP observations and using
models with $T_{\rm eff}$=70\,000\,K. The reasons for this more deviating
result can be i) the lower, probably more realistic $T_{\rm eff}$ of our
model (based on the analysis of \cite{haas:phd}), ii) our use of
self-consistent metal-line blanketed models in contrast to the Becker \&
Butler approach, and iii) an increased accuracy of our analysis using
combined ORFEUS\,\textsc{ii}, IUE-SWP and IUE-LWR observations.

At present the accuracy of Fe and Ni abundance determinations is limited
mainly by uncertainties of the atomic data available and by the S/N ratio
and resolution of the spectra. Therefore an error in the derived abundances
of a factor of about two has to be accepted. For a reliable analysis of
Far-UV spectra complementary Near-UV observations are recommended.
Uncertainties in the knowledge of interstellar reddening enlarges the
systematic error.

Generally, we have demonstrated the importance of correctly considerating
iron-group line blanketing for the analysis of future high-resolution
Far-UV spectra of hot compact stars. This more detailed analysis technique
enables to fully benefit from the improved high-resolution spectra of FUSE.

% --- acknowledgments ------------------------------------------------
%
\begin{acknowledgements}
  We would like to thank Stefan Dreizler, Stefan Haas, Thomas Rauch, and
  Klaus Werner for many helpful discussions and careful reading of the
  manuscript. This work is supported by the Deutsches Zentrum f\"ur Luft-
  und Raumfahrt (DLR) under grant 5O\,QV\,97054. The IUE data presented in
  this paper were obtained from the Multimission Archive at the Space
  Telescope Science Institute (MAST).
\end{acknowledgements}

% --- bibliography ---------------------------------------------------
%

% --- end document ---------------------------------------------------
%
\end{document}